\documentclass[a4paper,11pt]{article}
\pdfoutput=1 

\usepackage{jcappub} 
\usepackage[normalem]{ulem}      
\usepackage{bm}

\usepackage[T1]{fontenc} 
\usepackage{bigints}
\usepackage{longtable}
\usepackage[Symbol]{upgreek}

\usepackage[dvipsnames,table,xcdraw]{xcolor}
\newcommand{\be}{\begin{equation}}
\newcommand{\ee}{\end{equation}}
\newcommand{\beq}{\begin{equation}}
\newcommand{\eeq}{\end{equation}}
\usepackage{bm}
\newcommand{\de}{{\textrm d}}

\def\lsim{\mathrel{\raise.3ex\hbox{$<$\kern-.75em\lower1ex\hbox{$\sim$}}}}
\def\gsim{\mathrel{\raise.3ex\hbox{$>$\kern-.75em\lower1ex\hbox{$\sim$}}}}

\title{\boldmath Unraveling TeV Halos with the Cherenkov Telescope Array}

\author[a,b,c]{Dan Hooper,}
\author[a,b]{Elena Pinetti,}
\author[a,b]{Anastasia Sokolenko}

\affiliation[a]{Fermi National Accelerator Laboratory, Theoretical Astrophysics Department, Batavia, IL, 60510, USA}
\affiliation[b]{University of Chicago, Kavli Institute for Cosmological Physics, Chicago, IL 60637, USA}
\affiliation[c]{University of Chicago, Department of Astronomy \& Astrophysics, Chicago, IL 60637, USA}

\abstract{Pulsars are observed to emit bright and spatially extended emission at multi-TeV energies. Although such ``TeV halos'' appear to be an approximately universal feature of middle-aged pulsars, there remains much to be understood about these systems. In this paper, we project the ability of the Cherenkov Telescope Array (CTA) to measure the properties of TeV halos, focusing on the case of the nearby Geminga pulsar. We conclude that CTA will be able to provide important information about this source, allowing us to discriminate between a range of different models that are currently consistent with all existing data. In particular, such observations will help us to measure the normalization, energy dependence, and spatial dependence of the diffusion coefficient in the region that surrounds Geminga, as well as the spectrum of the electrons that are injected from this source.}

\begin{document}
\maketitle
\flushbottom

\section{Introduction}

In 2017, the High-Altitude Water Cherenkov (HAWC) observatory reported the detection of bright and spatially extended multi-TeV emission from the regions surrounding the Geminga and Monogem pulsars~\cite{HAWC:2017kbo, Abeysekara:2017hyn} (see also Ref.~\cite{Abdo:2009ku}). The spectrum and intensity of this emission reveal that these sources convert on the order of 10\% of their total spin-down power into very high-energy electron-positron pairs. Furthermore, as each of these ``TeV halos'' is observed to extend out to $\sim 5^{\circ}$ in radius (corresponding to approximately $\sim 25 \, {\rm pc}$), these observations indicate that cosmic rays propagate much less efficiently in the vicinity of these sources than they do elsewhere in the interstellar medium (ISM)~\cite{Hooper_2017,2005yCat.7245....0M,Lopez-Coto:2017pbk,Johannesson:2019jlk,DiMauro:2019hwn,Evoli:2018aza,Fang:2019iym,Hooper:2021kyp}. 

Over the past few years, data from HAWC, HESS, and LHAASO has been used to identify TeV halos around many other middle-aged ($t_{\rm age} \sim 10^5-10^6 \, {\rm yr}$) pulsars~\cite{Linden:2017vvb,Sudoh:2019lav,Sudoh:2021avj,HAWC:2020hrt,2019ICRC...36..797S,HESS:2018pbp,HESS:2017lee,LHAASO:2023rpg,LHAASO:2021crt,Fang:2021qon}, supporting the conclusion that TeV halos are an approximately universal feature of such objects. In contrast, younger pulsars, such as the Crab ($t_{\rm age} \approx 964 \, {\rm yr}$), have not been observed to produce extended multi-TeV emission~\cite{Aharonian_2006, Albert_2008, Abeysekara_2017}. These and other observations suggest that pulsars undergo several stages of evolution~\cite{Liu:2022hqf}. In the earliest of these stages ($t_{\rm age} \lsim 10^4 \, {\rm yr}$), electrons and positrons are confined within a so-called pulsar wind nebula. During this time, the powerful magnetic field of the neutron star accelerates charged particles which subsequently interact with the surrounding medium to create a termination shock. This shock leads to a second stage in which the shock fragments the pulsar wind nebula, allowing the cosmic rays to escape and propagate into the surrounding ISM. As an intermediate case, we note that the Vela pulsar ($t_{\rm age} \approx 11$ kyr) does not appear to have a typical TeV halo, but is surrounded by a $\sim 10$ pc region that produces significant emission in the GeV and radio bands~\cite{Abdo_2010, Abramowski_2012, Grondin_2013, Tibaldo_2018}. Vela could thus potentially represent an example of a pulsar that is in a transition between its pulsar wind nebula and TeV halo stages~\cite{Sudoh:2019lav}.

Young neutron stars also generate significant gamma-ray emission as supernova remnants. Whereas both pulsar wind nebulae and TeV halos are powered by a pulsar's rotational kinetic energy, supernova remnants rely on the energy that is liberated in a supernova explosion. In contrast to pulsar wind nebulae and TeV halos, supernova remnants grow steadily, at a rate that depends on the density and other characteristics of the surrounding ISM. Such objects ultimately become much larger than either pulsar wind nebulae or TeV halos, with radii that extend out to $\sim 50-100 \, {\rm pc}$~\cite{Xi:2018mii,Stafford:2018zoy}. Furthermore, while supernova remnants persist longer than pulsar wind nebulae, they are not nearly as long-lived as TeV halos. In particular, supernova remnants become faint as their shocks slow down, typically on a timescale of tens of thousands of years~\cite{Sarbadhicary:2016sch}.

As the very high-energy electrons and positrons accelerated by a pulsar diffuse away from their source and scatter with the radiation in the surrounding ISM, a TeV halo is formed. Geminga and Monogem are each prototypical examples of pulsars in this stage of evolution. The observed angular extent of TeV halos forces us to accept the puzzling fact that cosmic rays propagate in the vicinity of TeV halos much more slowly than they do elsewhere in the ISM~\cite{HAWC:2017kbo, Abeysekara:2017hyn, Kappl:2015bqa, Genolini:2019ewc}. Furthermore, the intensity of the multi-TeV gamma-ray emission from these objects implies that a significant fraction of their pulsars' total spin-down power is being converted into the acceleration of very high-energy electrons and positrons. Among other implications, this supports the conclusion that pulsars are responsible for generating the cosmic-ray positron excess, as reported by the PAMELA and AMS-02 collaborations~\cite{Hooper_2017,Profumo_2018,Fang_2018,DiMauro:2020cbn,Evoli:2020szd, Manconi:2020ipm,Orusa:2021tts,Bitter:2022uqj} (see also Refs.~\cite{Hooper:2008kg,Yuksel:2008rf,Abdo:2009ku,Malyshev:2009tw,DiMauro:2014iia, Boudaud:2014dta}).

Of the approximately 3400 pulsars that have been detected to date, the vast majority of these objects have been observed only at radio wavelengths~\cite{Manchester:2004bp, ATNF}. Such pulsars are characterised by their emission of pulsating electromagnetic radiation, which is visible only to an observer that is aligned along their magnetic axes. As a result, it is reasonable to conclude that most of the Milky Way's pulsars -- those with radio beams that are {\it not} pointed in our direction -- have not yet been detected. In contrast to their radio beams, the very high-energy emission associated with a pulsar's TeV halo is emitted isotropically. TeV halos thus represent a powerful means by which to discover pulsars whose radio beams are not aligned in our direction~\cite{Linden:2017vvb}. 

There remains much to understand about the physics of TeV halos. In particular, it is not yet known why or how the process of diffusion is inhibited in the volume surrounding these sources (for discussions, see Refs.~\cite{Mukhopadhyay:2021dyh,DeLaTorreLuque:2022chz,Liu:2019zyj,Recchia:2021kty,Bao:2021hey,Evoli:2018aza,Fang:2019iym}). To discriminate between various models that could potentially account for this observed behavior, it will be essential for us to measure the spectrum and angular distribution of the gamma-ray emission from TeV halos in much greater detail. Particularly promising in this regard is the upcoming Cherenkov Telescope Array (CTA), which will offer unprecedented angular resolution and overall sensitivity to gamma rays in the energy range of $E_{\gamma} \sim 10^2-10^5 \, {\rm GeV}$. In this paper, we consider the ability of CTA to distinguish between different models of TeV halos, focusing on the specific case of Geminga. To this end, we consider a variety of models with different values for the parameters associated with the injected electron spectrum, the time evolution of the pulsar's spin-down, and the diffusion coefficient surrounding the pulsar. We identify a variety of models that are currently consistent with all existing data, but that could be differentiated by CTA.\footnote{For a complementary study assessing the ability of CTA to study the characteristics of the Milky Way's TeV halo population, see Ref.~\cite{Eckner:2022gil}.}

\section{TeV-Scale Gamma Rays From Pulsars}

The propagation and energy losses of electrons\footnote{Throughout this paper, we will often refer to electrons and positrons simply as ``electrons''.} can be described by the following transport equation:
\begin{equation}
   \dfrac{\partial }{\partial t} \dfrac{\de n_e}{\de E_e} (E_e, \vec{r}, t) - \vec{\nabla} \cdot \left[D(E_e, \Vec{r}) \;  \Vec{\nabla} \, \dfrac{\de n_e}{\de E_e}(E_e, \Vec{r}, t) \right] + \dfrac{\partial}{\partial E_e} \left[b_{\rm tot}(E_e) \dfrac{\de n_e}{\de E_e}(E_e, \Vec{r}, t) \right] = Q(E_e, \Vec{r}, t) \; , \nonumber \\
\end{equation}
where $d n_e/dE_e$ is the differential number density of electrons, $D$ is the diffusion coefficient, and $b_{\rm tot}$ is the total energy loss rate from inverse Compton scattering and synchrotron emission. On the right-hand side, $Q$ is the injected spectrum, representing the source of the electrons in question.

We model a pulsar as a point source of energetic electrons and assume that the injected spectrum follows a power-law with an exponential cut-off \cite{Delahaye:2010ji}, allowing us to write the source terms as follows:
\begin{equation}
    Q(E_e, \Vec{r}, t) = Q_\star \, E_e^{-\alpha} \, \exp \left(- \dfrac{E_e}{E_{\rm cut}} \right) \, L(t) \, \delta(\vec{r}) \; ,
\end{equation}
where $Q_{\star}$ is the normalization of the spectrum, $\alpha$ and $E_{\rm cut}$ are spectral index and cutoff of the injected electrons, and $L(t)$ accounts for the time dependence of the luminosity in injected electrons.
We take the electron luminosity to be proportional to the rate at which the pulsar loses rotational kinetic energy (i.e., its spin-down power), which can be expressed as
\begin{equation}
    L(t) = L_0 \left(1 + \dfrac{t}{\tau} \right)^{-\dfrac{n+1}{n-1}} \; ,
\end{equation}
where $L_0$ is the initial spin-down luminosity, $n$ the braking index, and $\tau$ is the characteristic spin-down timescale. While many pulsars exhibit breaking indices near $n = 3$ (corresponding to the case of magnetic dipole breaking), others evolve as rapidly as $n \sim 1.4$~\cite{Kaspi:2002pu}. We further introduce the quantity $\eta$, which is the fraction of a pulsar's total spin-down power that goes into the production of electron-positron pairs with $E_e > 0.1 \, {\rm GeV}$, $L=\eta \dot{E}_{\rm rot}$.

In the energy range of interest, electrons lose energy through a combination of inverse Compton scattering and synchrotron radiation,
$b_{\rm tot} = b_{\rm sync} + b_{\rm ICS}$~\cite{Cirelli:2010xx}. These contributions to the energy loss rate are given by
\begin{align}
b_{\rm sync} &= \frac{2 \sigma_{\rm t} c B^2}{3 m^2_e \mu_0} \bigg(\frac{E_e}{m_e}\bigg)^2 \, , \\
   b_{\rm ICS} &= \sum_i  \frac{4\sigma_{\rm t} u_i \, S_i(E_e)}{3c^3}   \bigg(\frac{E_e}{m_e}\bigg)^2 \, ,
\end{align}
where $\sigma_{\rm t}$ is the Thomson cross-section, $\mu_0$ is the permeability constant, and we take the strength of the magnetic field to be $B = 3 \, \mu G$.
The sum in the expression for the inverse Compton losses runs over the different components of the interstellar radiation field, which we take to be the cosmic microwave background ($T_{\rm CMB} = 2.75$~K, $u_{\rm CMB} = 0.26$ eV/cm$^3$), infrared emission from dust ($T_{\rm IR} = 20$~K, $u_{\rm IR} = 0.60$ eV/cm$^3$), and optical starlight ($T_{\rm SL} = 5000$~K, $u_{\rm SL} = 0.60$ eV/cm$^3$)~\cite{Cirelli:2009vg, Vernetto:2016alq, Hooper_2017}. At very high energies, $E_e > {m_e^2}/{2 T}$, inverse Compton scattering occurs in the Klein-Nishina regime, characterized by the following suppression factor:
\begin{equation}
    S_i (E_e) = \dfrac{A_i}{A_i + \left({E_e}/{m_e}\right)^2} \; ,
\end{equation}
where $A_i = 45 m_e^2/64 \pi^2 T_i^2$. 

For the diffusion coefficient, we consider in this study a two-zone model, in which we take the magnitude and energy dependence of the diffusion coefficient to change at a distance, $r_h$, from the pulsar,
\begin{equation}
    D(E_e) = \Bigg\{ \begin{array}{ll} 
    D_0 \left(E_e/{1 \,\rm GeV} \right)^\delta \quad &r \leq r_h \\[3pt]
    D_{\rm ISM} \left(E_e/{1 \,\rm GeV} \right)^{\delta_{\rm ISM}} \quad &r > r_h \; . \\
    \end{array}
\end{equation}

The observed angular extent of TeV halos strongly favors two-zone models over those with a uniform diffusion coefficient. In particular, whereas diffusion in the ISM is characterized by $D_{\rm ISM} \sim 4 \times 10^{28} \, {\rm cm}^2/{\rm s}$ and $\delta_{\rm ISM} \sim 1/3$~\cite{Strong:2007nh,Trotta:2010mx}, the observed morphology of the Geminga and Monogem halos each require $D_0 \sim 10^{26} \, {\rm cm}^2/{\rm s}$ (for $\delta =1/3$)~\cite{Hooper_2017, Fang_2018, Profumo_2018, Tang_2019, J_hannesson_2019}.

The diffusion coefficient and energy loss rates can be used together to determine the diffusion length, $\lambda$, over which electrons are typically displaced:
\begin{equation}
    \lambda(E_e, E_0) = \Bigg[ 4\int_{E_0}^{E_e} \, \de E_e' \; \dfrac{D(E_e')}{b_{\rm tot}(E_e')} \Bigg]^{1/2} ,
\end{equation}
where $E_0$ and $E_e$ are the initial and final electron energies, respectively. We can further relate these energies to the amount of time that has passed since the electrons were injected:
\begin{equation}
    \int_{E_0}^{E_e} \dfrac{1}{b_{\rm tot}(E_e')} = t_\star \; ,
    \label{E0}
\end{equation}
where $t_{\star} \equiv t_{\rm obs} - t_{\rm inj}$ is the difference between the time of the observation and time at which the electrons were injected from the pulsar. This allows us to treat the diffusion length as a function of the final energy and the time since injection, $\lambda(E_e,t_{\star})$.
Note that if the diffusion coefficient in the region surrounding Geminga had been similar to that observed elsewhere in the ISM, the inverse Compton emission due to 35 TeV electrons, would extend out to a distance of $\lambda \sim 200 \, {\rm pc}$, 
corresponding to an angular scale of $\sim 60^{\circ}$, far beyond the $\sim 2^{\circ}$ extension observed by HAWC and Milagro. It is this key observation that forces us to conclude that diffusion is very inefficient in the vicinity of these pulsars.

The distribution of the energetic electrons from a TeV halo can be expressed as
\begin{align}
\dfrac{\de n_e}{\de E_e} (E_e, r, t) = \int^t_0 \frac{dt' \, Q_{\star} \, L(t') \, E_0 \, (E_e,t-t')^{2-\alpha}}{\pi^{3/2} \, E^2_e \, \lambda(E_e, t-t')^3} \exp\bigg[-\frac{E_0(E_e,t-t')}{E_{\rm cut}}\bigg]  \exp\bigg[-\bigg(\frac{r}{\lambda(E_e,t-t')}\bigg)^2\bigg], \nonumber \\
\end{align}
where $E_0$ is evaluated using Eq.~\eqref{E0} with $t_\star = t - t'$. Keep in mind that both $E_0$ and $\lambda$ are functions of $E_e$.

The second element that we will need in order to compute the gamma-ray flux from a TeV halo is the spectrum of inverse Compton emission that is produced by a given high-energy electron. The differential spectrum of inverse Compton emission radiated from an electron of energy $E_e$ is given by~\cite{Blumenthal:1970gc,Delahaye:2010ji,Cirelli:2010xx}
\begin{align}
\frac{dN_{\gamma}}{dE_{\gamma}}(E_{\gamma}, E_e) = c \int d\epsilon \frac{dn}{d\epsilon}(\epsilon) \,  \frac{d\sigma_{\rm ICS}}{dE_{\gamma}}(E_{\gamma}, \epsilon,E_e), 
\end{align}
where $d\sigma_{\rm ICS}/dE_{\gamma}$ is the differential cross section for inverse Compton scattering~\cite{1981Ap&SS..79..321A} and $dn/d\epsilon$ is the differential number density of target radiation. We take this radiation to consist of a sum of blackbodies associated with the cosmic microwave background, infrared emission from dust, and starlight, with energy densities and temperatures as described earlier in this section.

To obtain the flux of photons that reach Earth, we convolve the electron density with the spectrum of inverse Compton emission per electron,
\begin{equation}
    \dfrac{d \phi_{\gamma}}{d E_\gamma}\left(E_\gamma \right) = \int d \phi \int d \theta \sin{\theta} \int_{los} d s \int d E_e \, \dfrac{d n_e}{d E_e} (E_e, \Vec{r}\,) \, \dfrac{d N_\gamma}{d E_\gamma}(E_\gamma, E_e), \;
\end{equation}
where $s$ is the path along the line-of-sight ($los$).

\begin{figure}
    \centering
    \includegraphics[width=0.99\linewidth]{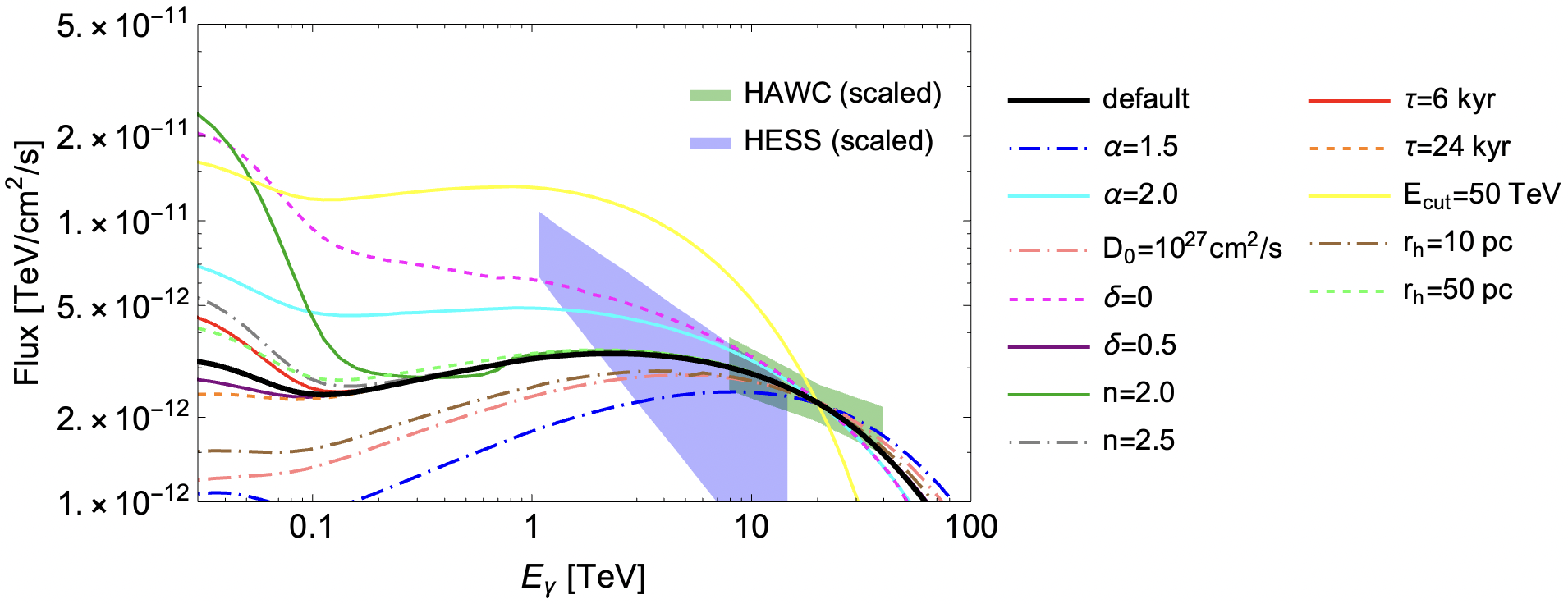}
    \caption{The impact of various parameters on the spectrum of the gamma-ray emission from a Geminga-like TeV halo, as integrated within a $2.5^{\circ}$ radius. These predictions are compared to Geminga's spectrum, as measured by HAWC~\cite{HAWC:2017kbo} and HESS~\cite{HESS:2023sbf}. For our default parameters, we have adopted $\alpha=1.8$, $D_0=10^{26} \, {\rm cm}^2/{\rm s}$, $\delta=0.33$, $n=3.0$, $r_h=30 \, {\rm pc}$, $\tau= 12 \, {\rm kyr}$, $E_{\rm cut}=500 \, {\rm TeV}$, and $\eta=0.25$. Each curve is normalized to the measured flux at $E_{\gamma}=20 \, {\rm TeV}$.}
    \label{fig:flux}
\end{figure}

\begin{figure}
    \centering
    \includegraphics[width=0.48\linewidth]{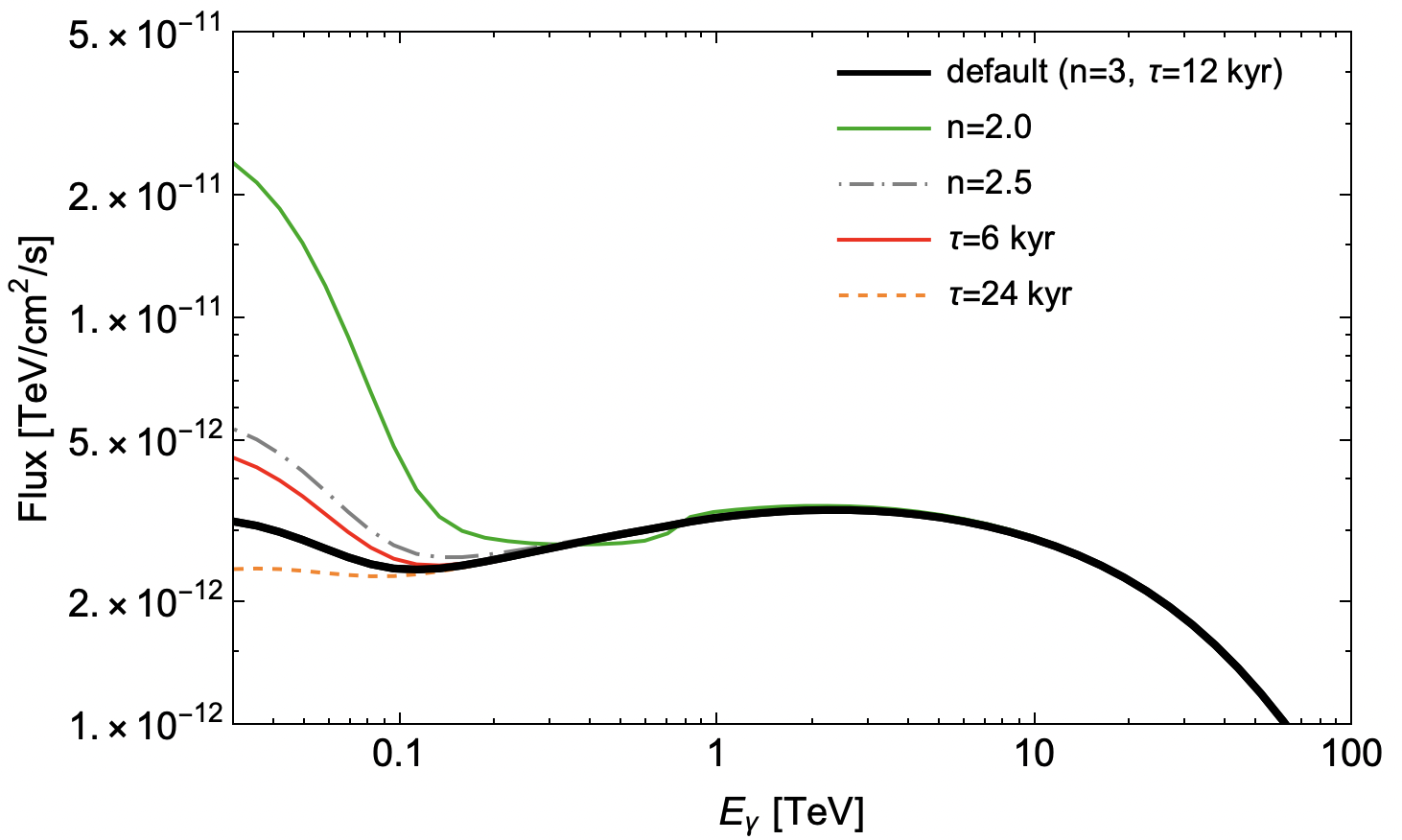}~\includegraphics[width=0.48\linewidth]{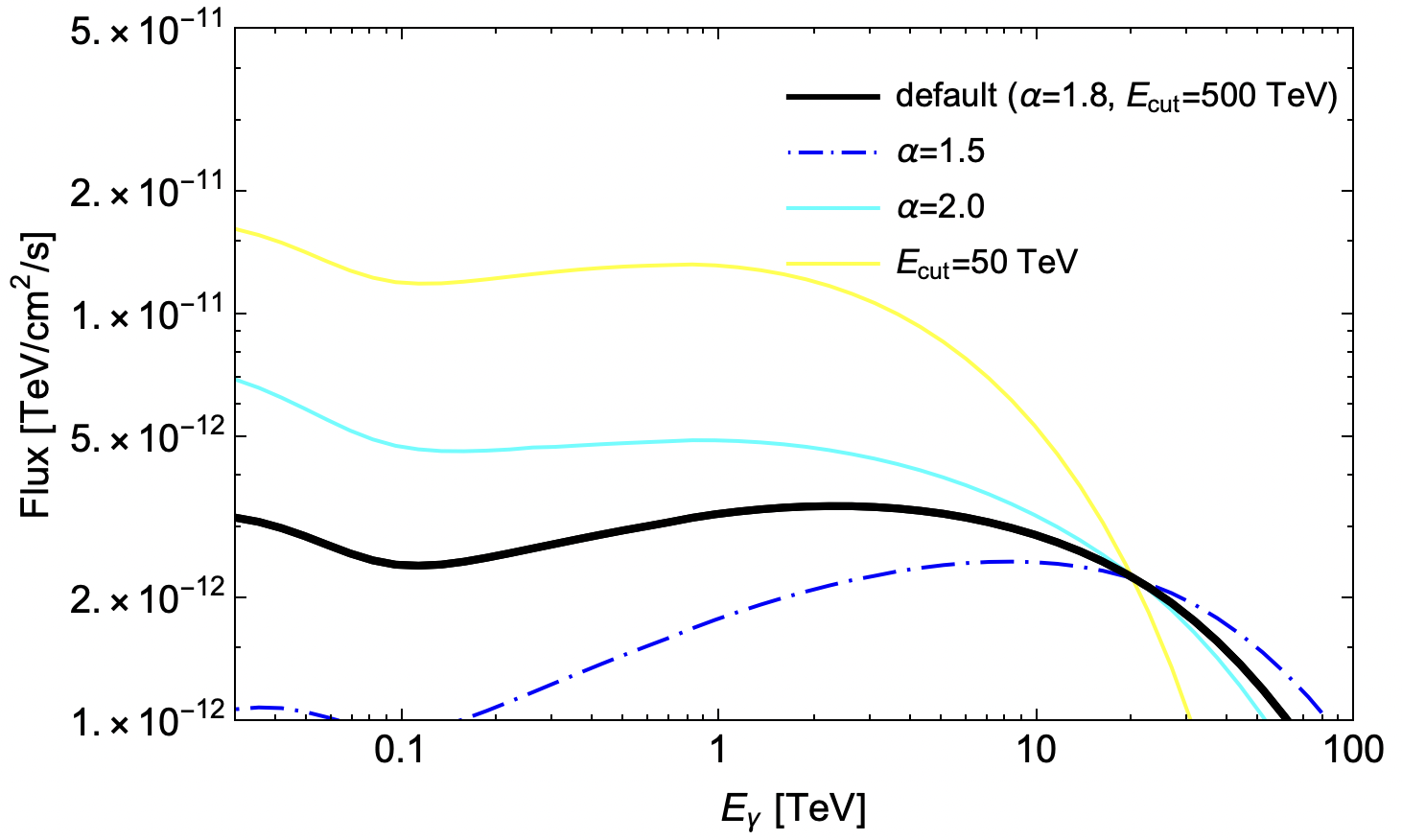}\\
    \includegraphics[width=0.48\linewidth]{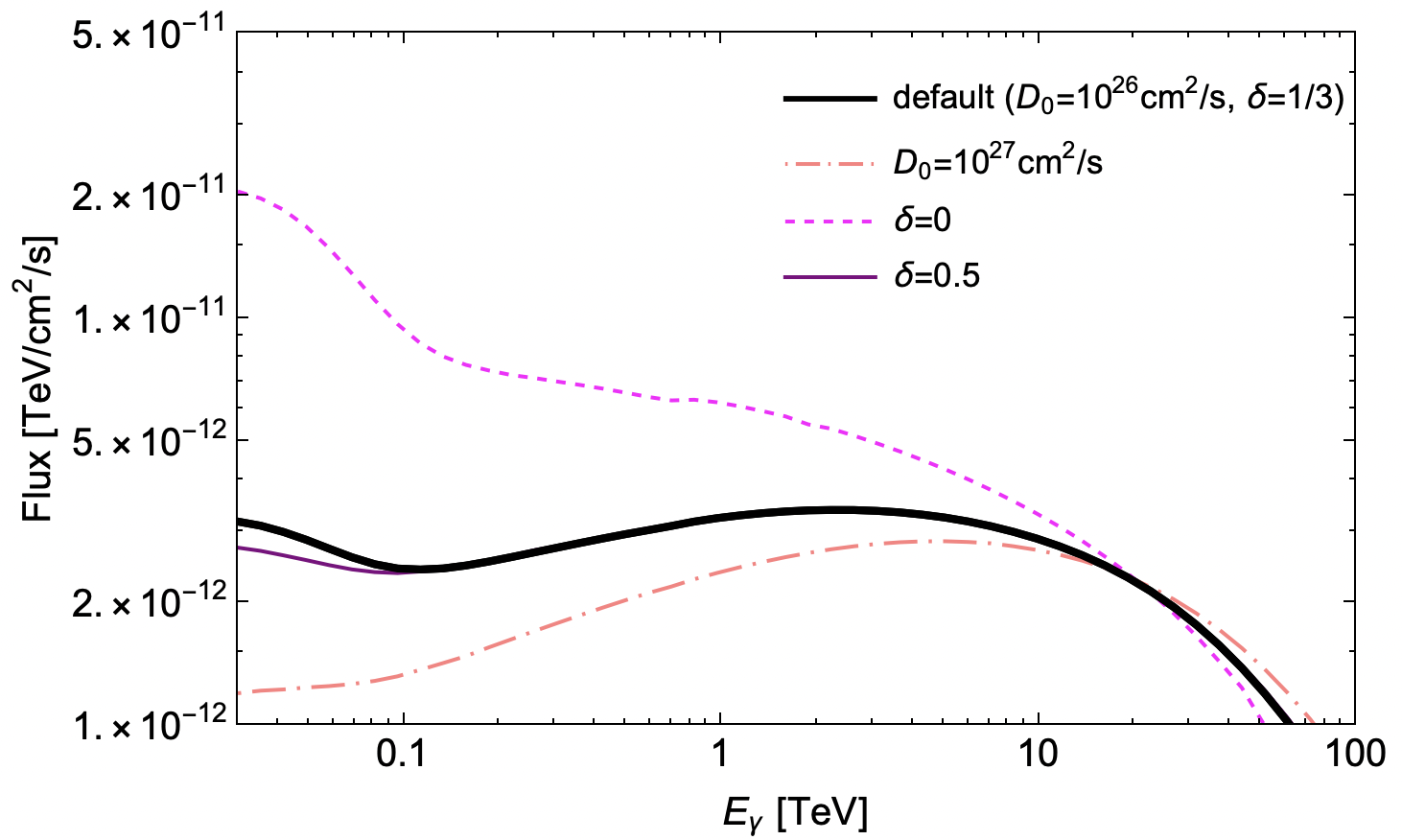}~\includegraphics[width=0.48\linewidth]{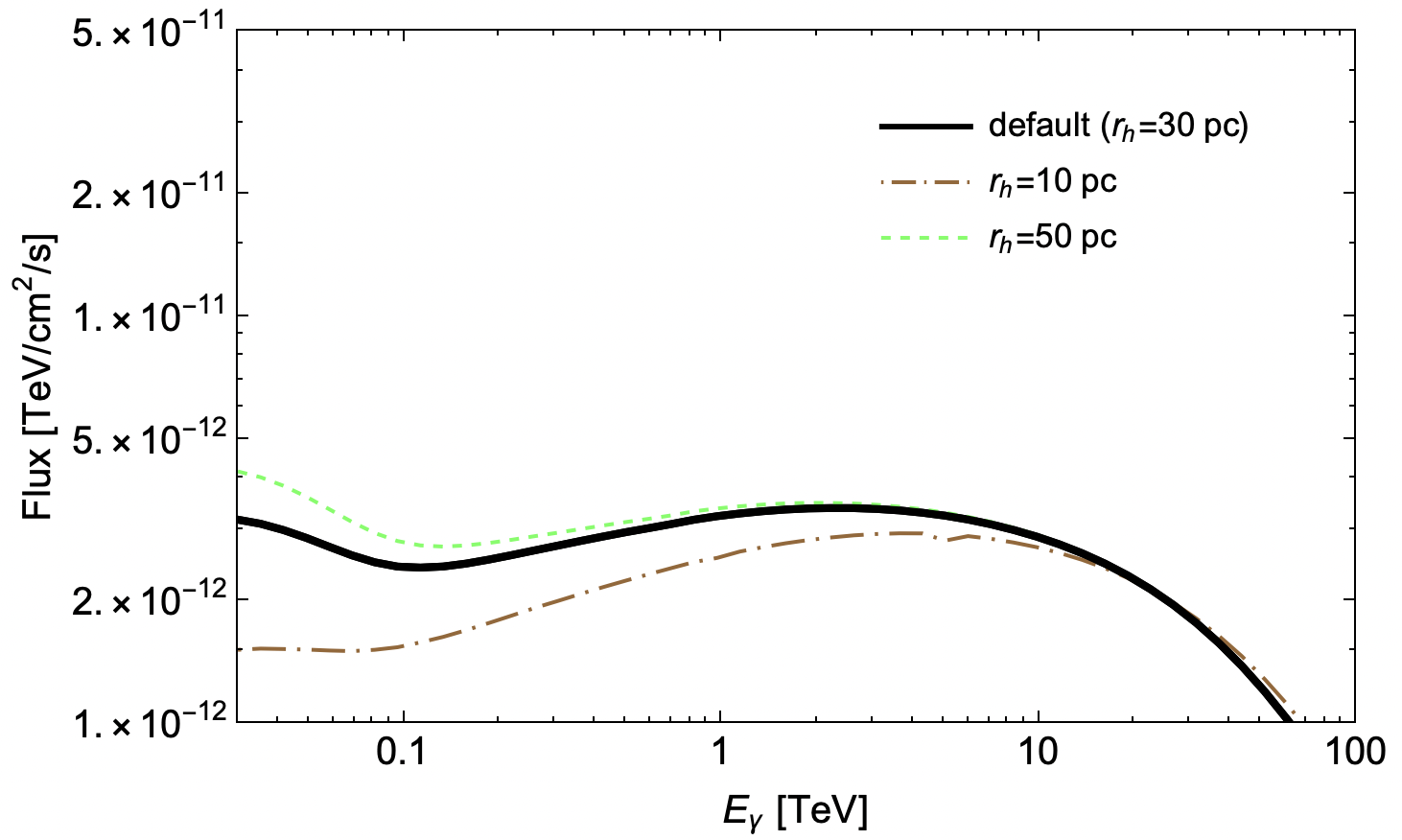}
    \caption{As in Fig.~\ref{fig:flux}, but showing the impact of different parameters in separate frames for clarity.}
    \label{fig:flux_3}
\end{figure}

In Figs.~\ref{fig:flux} and~\ref{fig:flux_3}, we show the impact of the various parameters described in this section on the spectrum of the gamma-ray emission from a Geminga-like TeV halo. For our default parameters, we have adopted $\alpha=1.8$, $D_0=10^{26} \, {\rm cm}^2/{\rm s}$, $\delta=0.33$, $n=3.0$, $r_h=30 \, {\rm pc}$, $\tau= 12 \, {\rm kyr}$, $E_{\rm cut}=500 \, {\rm TeV}$, and $\eta=0.25$. Each curve in this figure (and throughout this paper) is normalized such that it has the same flux at $E_{\gamma}=20 \, {\rm TeV}$ (see Table~\ref{tab:table} for the required efficiencies, $\eta$). We take the distance to Geminga to be 250~pc and its age to be 340~kyr.

In the upper left frame of Fig.~\ref{fig:flux_3}, we illustrate how the pulsar braking index, $n$, and the pulsar's spin-down timescale, $\tau$, each impact the shape of the gamma-ray spectrum. At high energies ($E_{\gamma} \gsim 0.3 \, {\rm TeV}$), these parameters do not significant impact the spectrum. This is because photons in this energy range are produced by very high-energy electrons which lose energy on a timescale that is much shorter than the age of the pulsar or the time that would be required for those particles to escape from the halo. The observed gamma-ray spectrum thus reflects the current injection rate of very high-energy electrons. At lower energies, in contrast, the values of $n$ and $\tau$ can each significantly impact the predicted spectrum. 
\begin{table}[t]
 \label{tab:table}
  \begin{center}
    \begin{tabular}{|c|c|} 
    \hline
      Model Name &   Efficiency, $\eta$\\
      \hline
      \hline
      Default & 28\% \\
      \hline
      $n=2.0$ & 28\%  \\
      \hline
      $n=2.5$ & 28\% \\
      \hline
      $\tau=6 \, {\rm kyr}$ & 28\% \\           
      \hline
      $\tau=24 \, {\rm kyr}$ & 28\% \\    
      \hline
      $D_0=10^{27} \, {\rm cm}^2/{\rm s}$ & 130\% \\
      \hline
      $\delta=0$   & 14\% \\
      \hline
      $\delta=0.5$ & 94\% \\     
      \hline      
      $\alpha=1.5$ & 17\% \\
      \hline
      $\alpha=2.0$ & 71\% \\  
      \hline
      $E_{\rm cut}=50 \, {\rm TeV}$ & 97\% \\     
      \hline
      $r_h=10 \, {\rm pc}$ & 43\% \\
      \hline
      $r_h=50 \, {\rm pc}$ & 28\% \\
      \hline
    \end{tabular}
    \caption{The efficiencies, $\eta$, adopted for TeV halo models used in this study. Our default model corresponds to $n=3.0$, $\tau= 12 \, {\rm kyr}$, $\alpha=1.8$, $E_{\rm cut}=500 \, {\rm TeV}$, $D_0=10^{26} \, {\rm cm}^2/{\rm s}$, $\delta=0.33$, and $r_h=30 \, {\rm pc}$.}
  \end{center}
\end{table}

In the lower left frame of Fig.~\ref{fig:flux_3}, we show how the diffusion parameters, $D_0$ and $\delta$, impact the predicted spectrum. Again, these parameters do not significantly impact the spectrum at the highest energies ($E_{\gamma} \gsim 10 \, {\rm TeV}$), but do at lower energies, where the timescale for electrons to escape the halo becomes comparable to the age of the pulsar. Note that the $D_0=10^{27} \, {\rm cm}^2/{\rm s}$ case requires an unphysical value of the efficiency, $\eta=1.3$, and is only shown for illustration. In the upper right frame of the same figure, we show the impact of changes to the spectrum of injected electrons, as parameterized by $\alpha$ and $E_{\rm cut}$. Lastly, in the lower right frame of this figure, we show the impact of the radius of the TeV halo, $r_h$ (beyond which the diffusion coefficient takes on standard ISM values).

\begin{figure}
    \centering\includegraphics[width=0.99\linewidth]{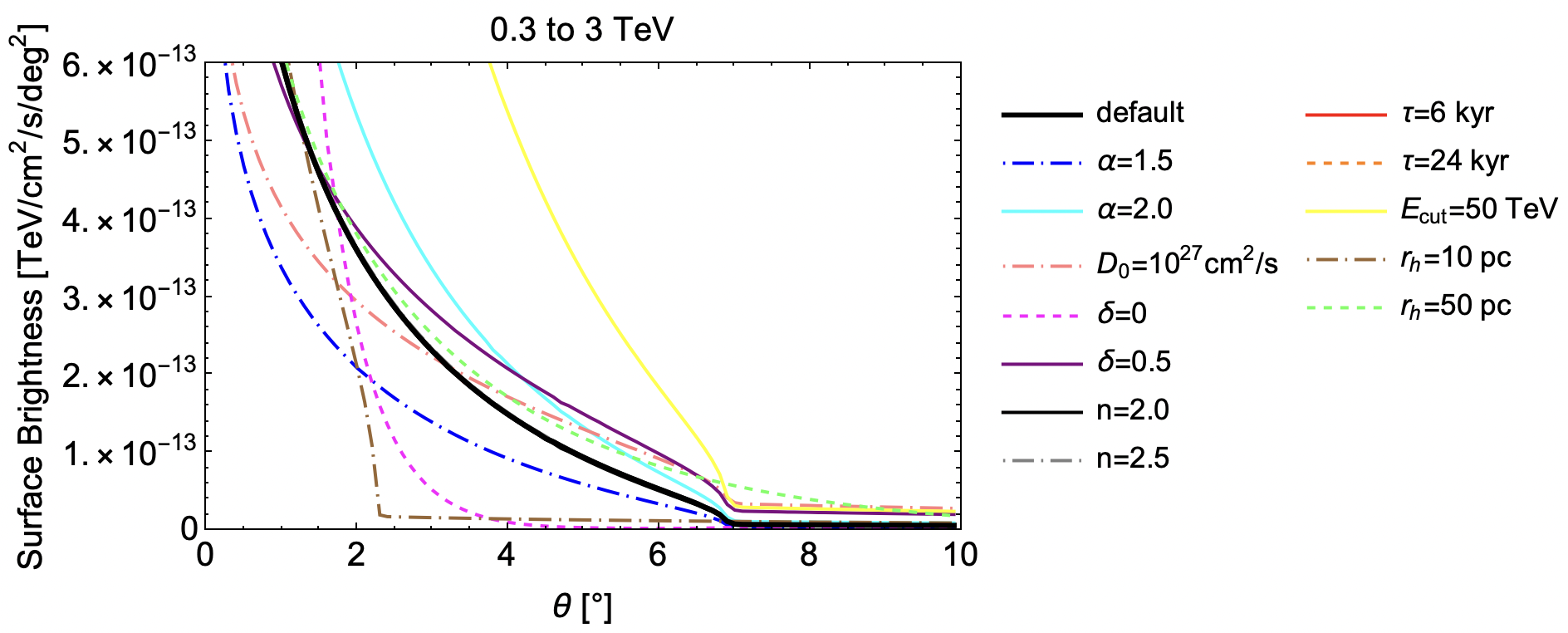}\\
    \includegraphics[width=0.99\linewidth]{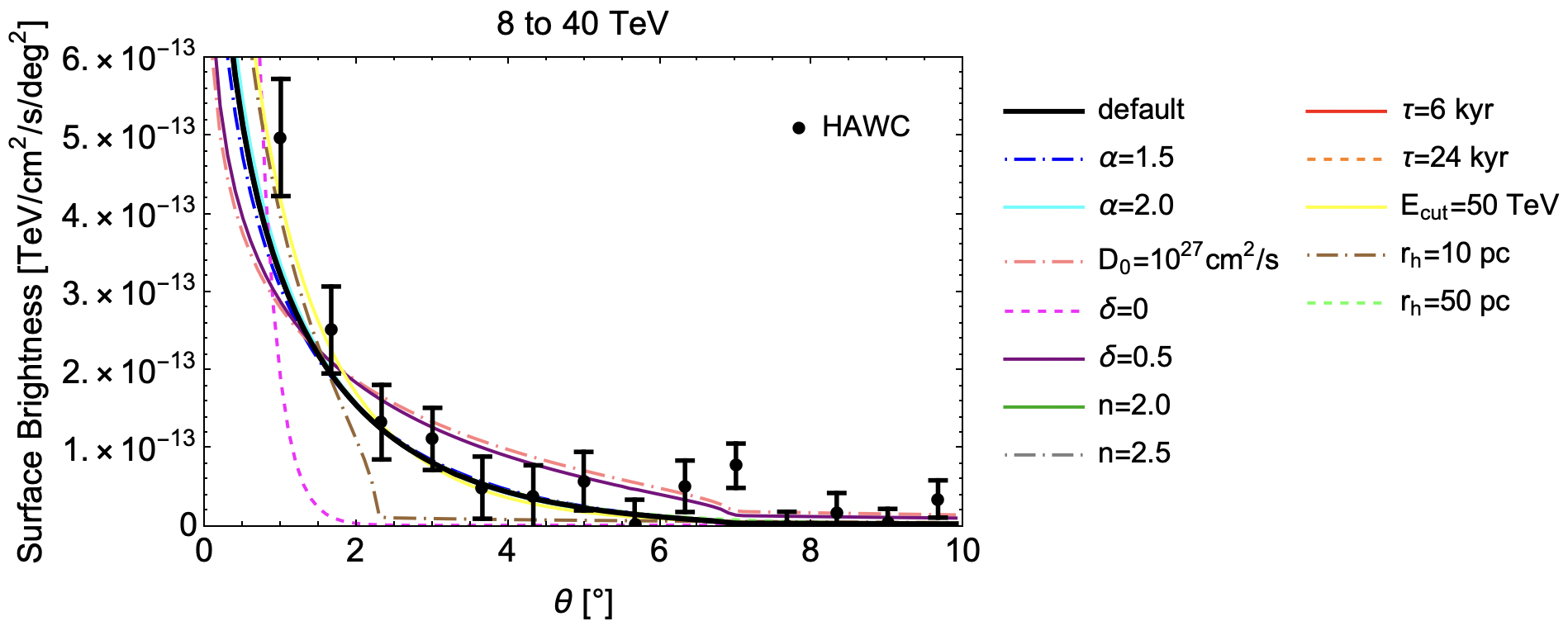}
    \caption{The impact of various parameters on the surface brightness profile of the gamma-ray emission from a Geminga-like TeV halo, integrated over two ranges of energy. The predictions for the 8-40 TeV case are compared in the lower frame to the measurements of HAWC~\cite{HAWC:2017kbo}. For our default parameters, we have adopted $\alpha=1.8$, $D_0=10^{26} \, {\rm cm}^2/{\rm s}$, $\delta=0.33$, $n=3.0$, $r_h=30 \, {\rm pc}$, $\tau= 12 \, {\rm kyr}$, $E_{\rm cut}=500 \, {\rm TeV}$, and $\eta=0.25$. Each curve is normalized such that it has the measured flux at $E_{\gamma}=20 \, {\rm TeV}$.}
    \label{fig:SB}
\end{figure}

\begin{figure}
    \centering
    \includegraphics[width=0.48\linewidth]{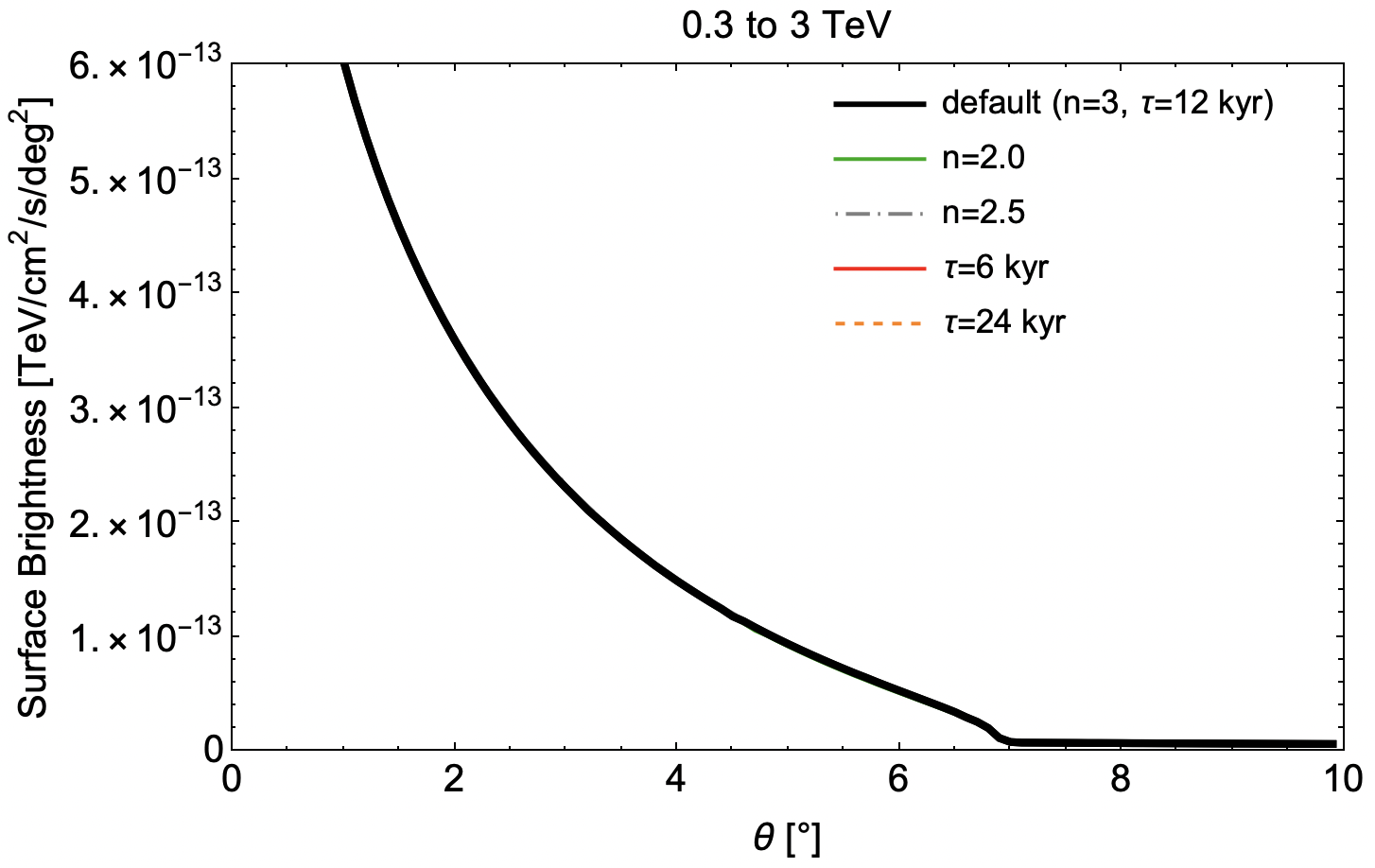}~\includegraphics[width=0.48\linewidth]{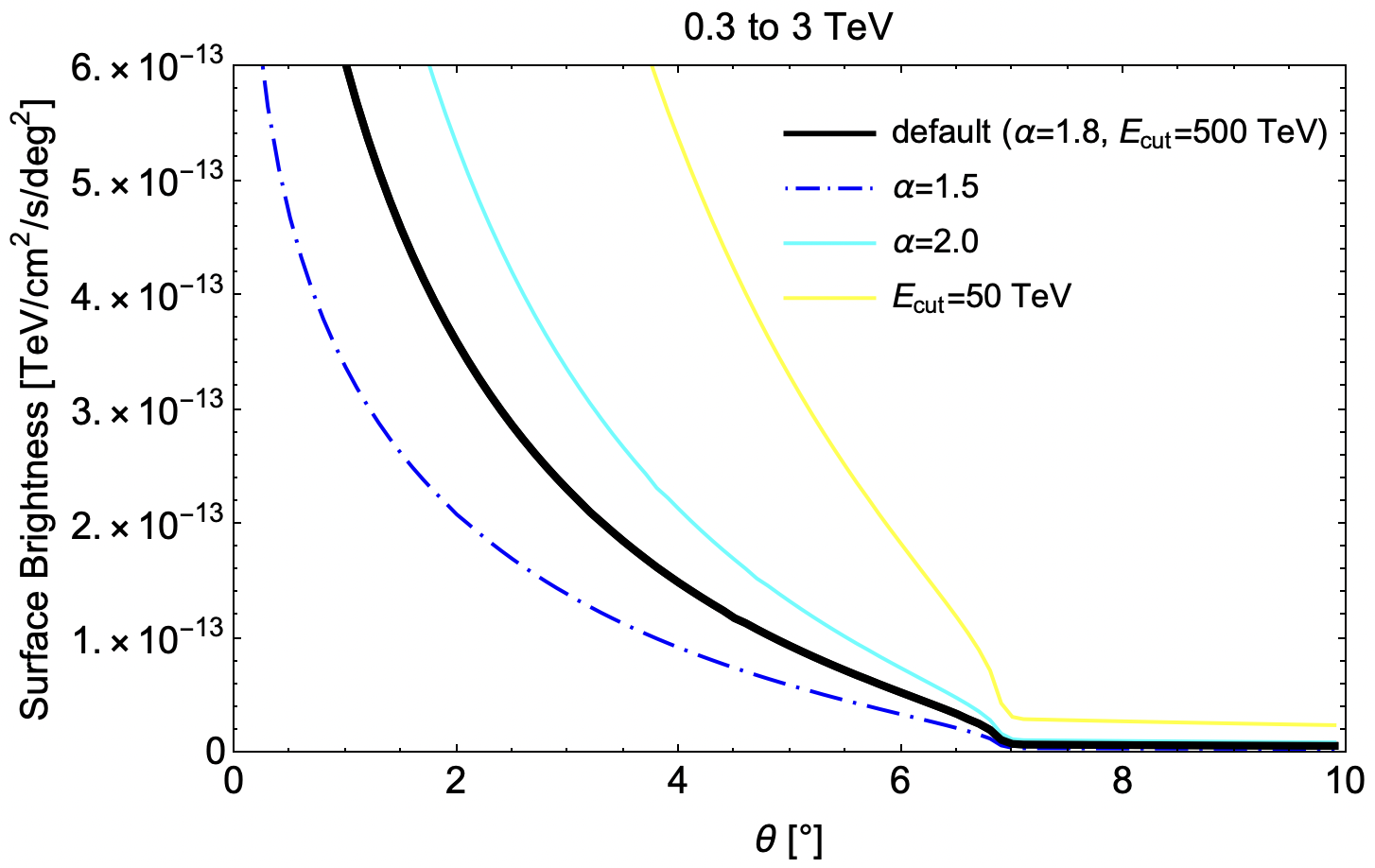}\\
    \includegraphics[width=0.48\linewidth]{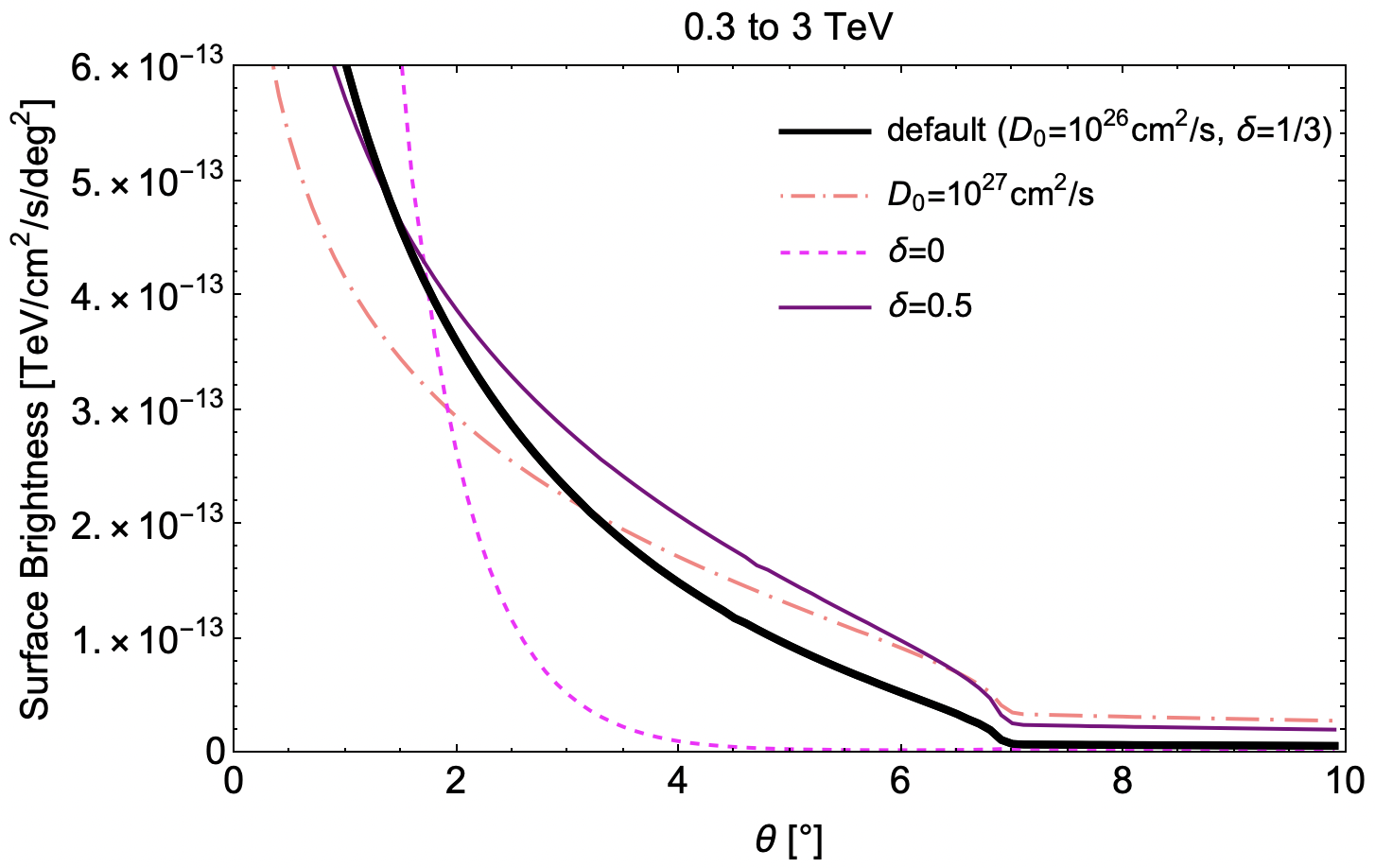}~\includegraphics[width=0.48\linewidth]{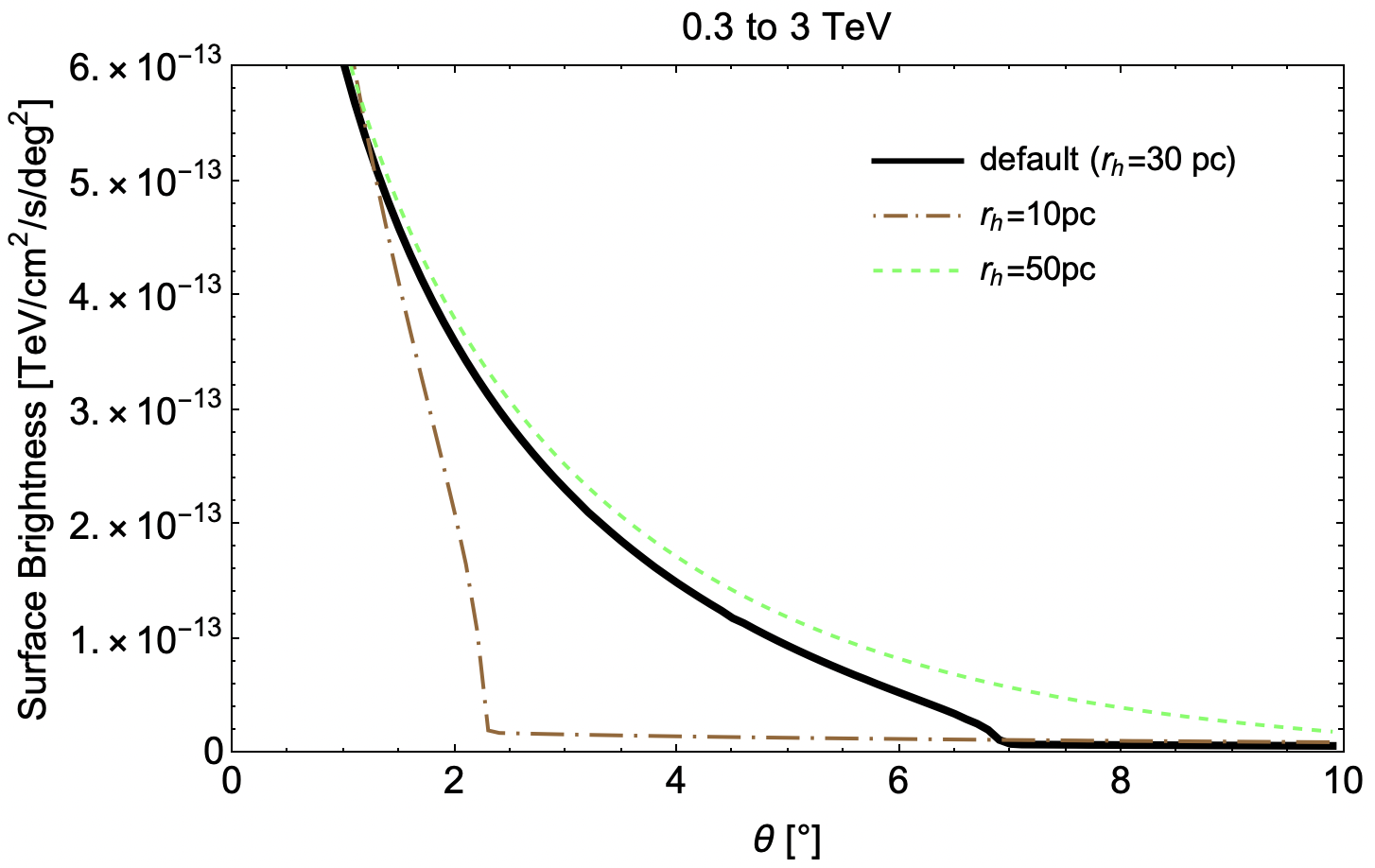}
    \caption{As in Fig.~\ref{fig:SB}, but showing the impact of different parameters in separate frames for clarity.}
    \label{fig:SB_2}
\end{figure}

\begin{figure}
    \centering
    \includegraphics[width=0.48\linewidth]{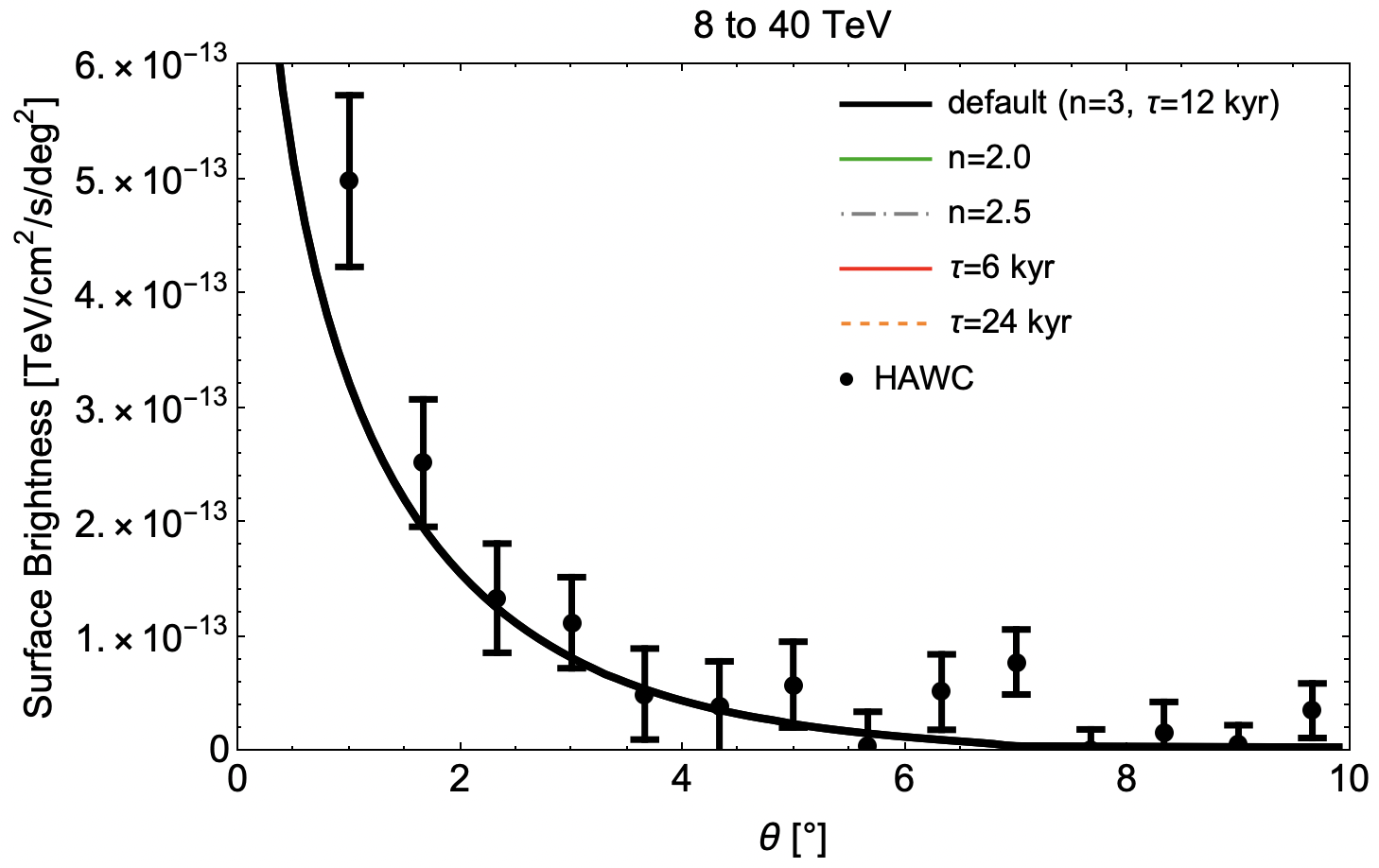}~\includegraphics[width=0.48\linewidth]{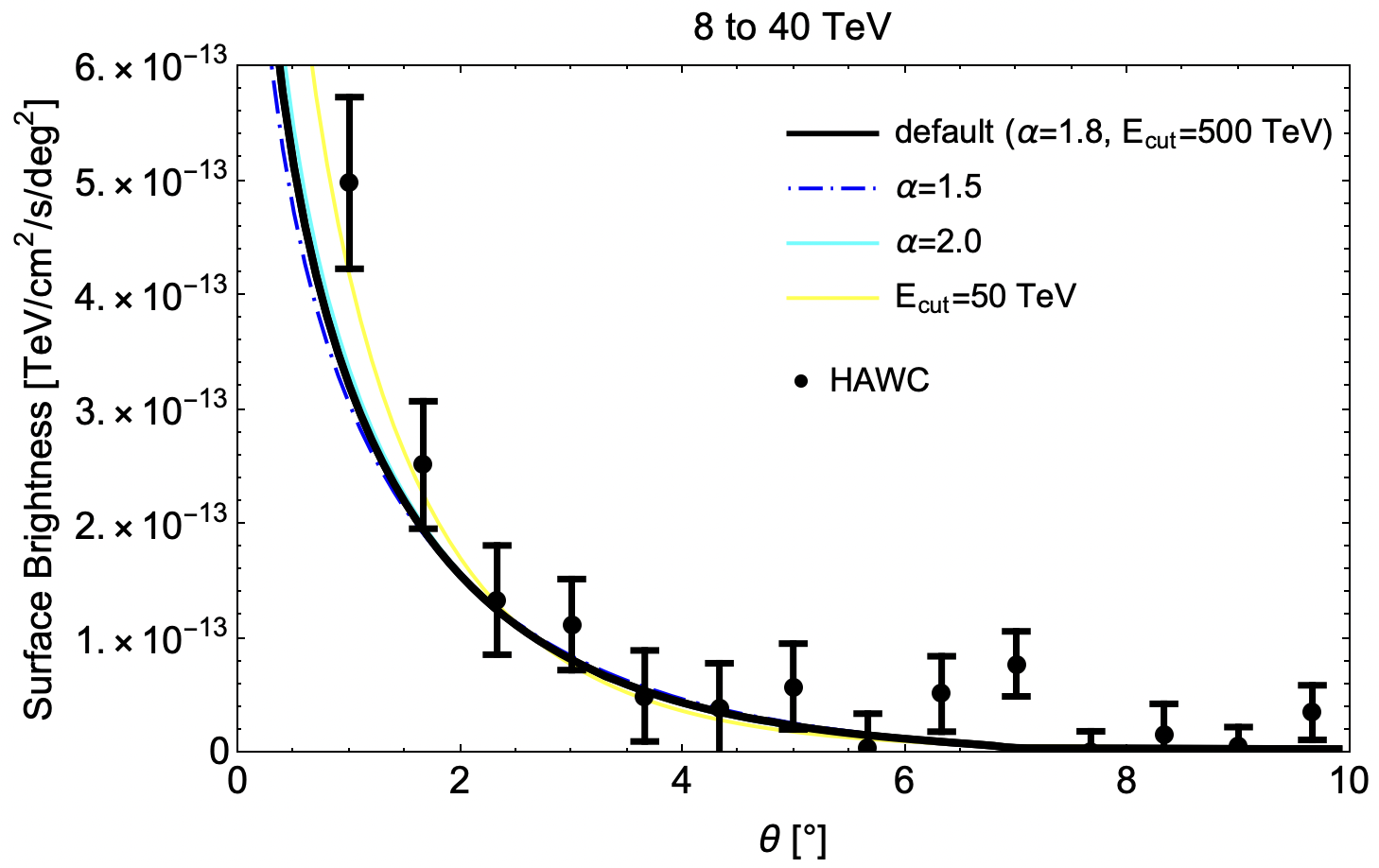}\\
    \includegraphics[width=0.48\linewidth]{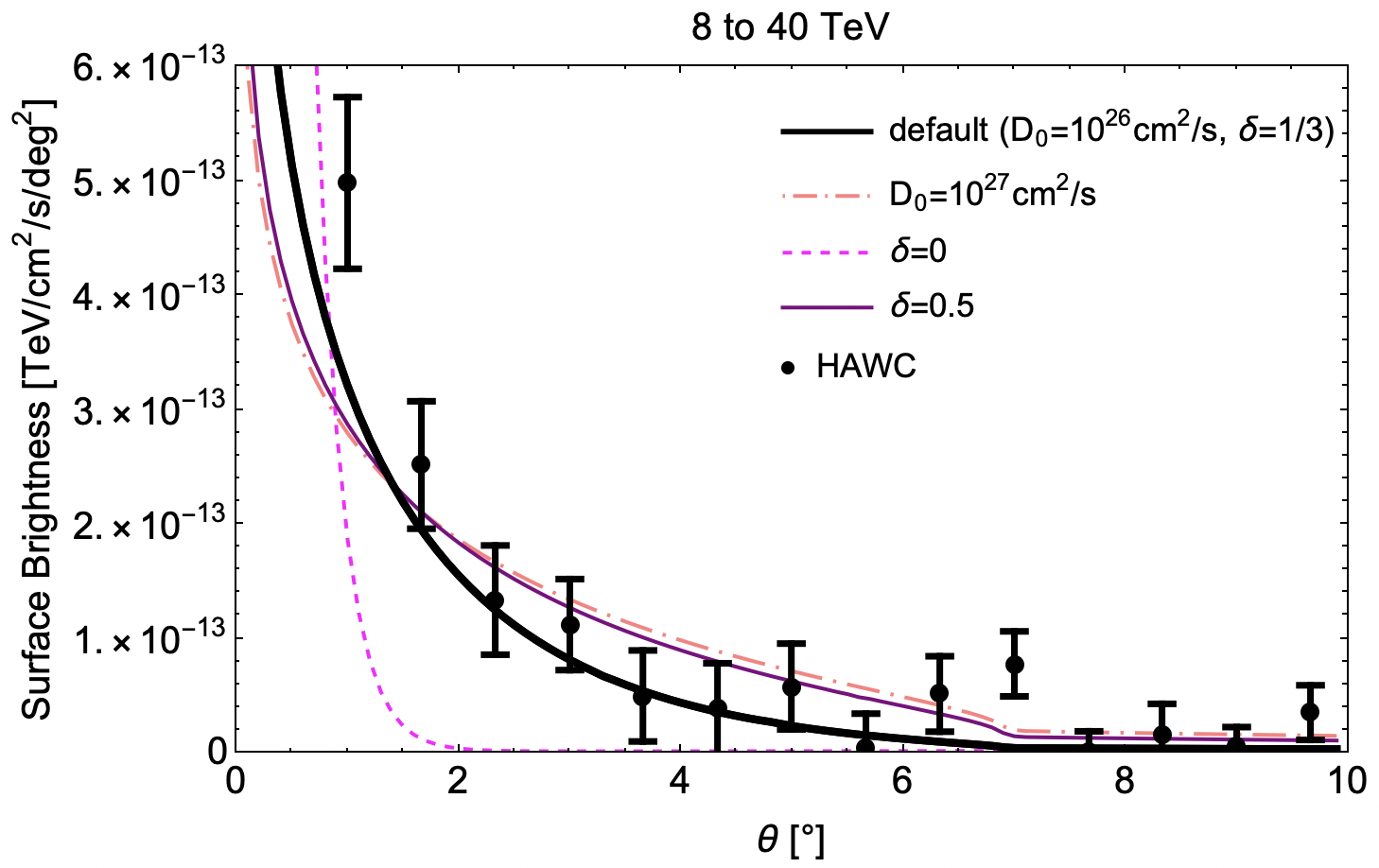}~\includegraphics[width=0.48\linewidth]{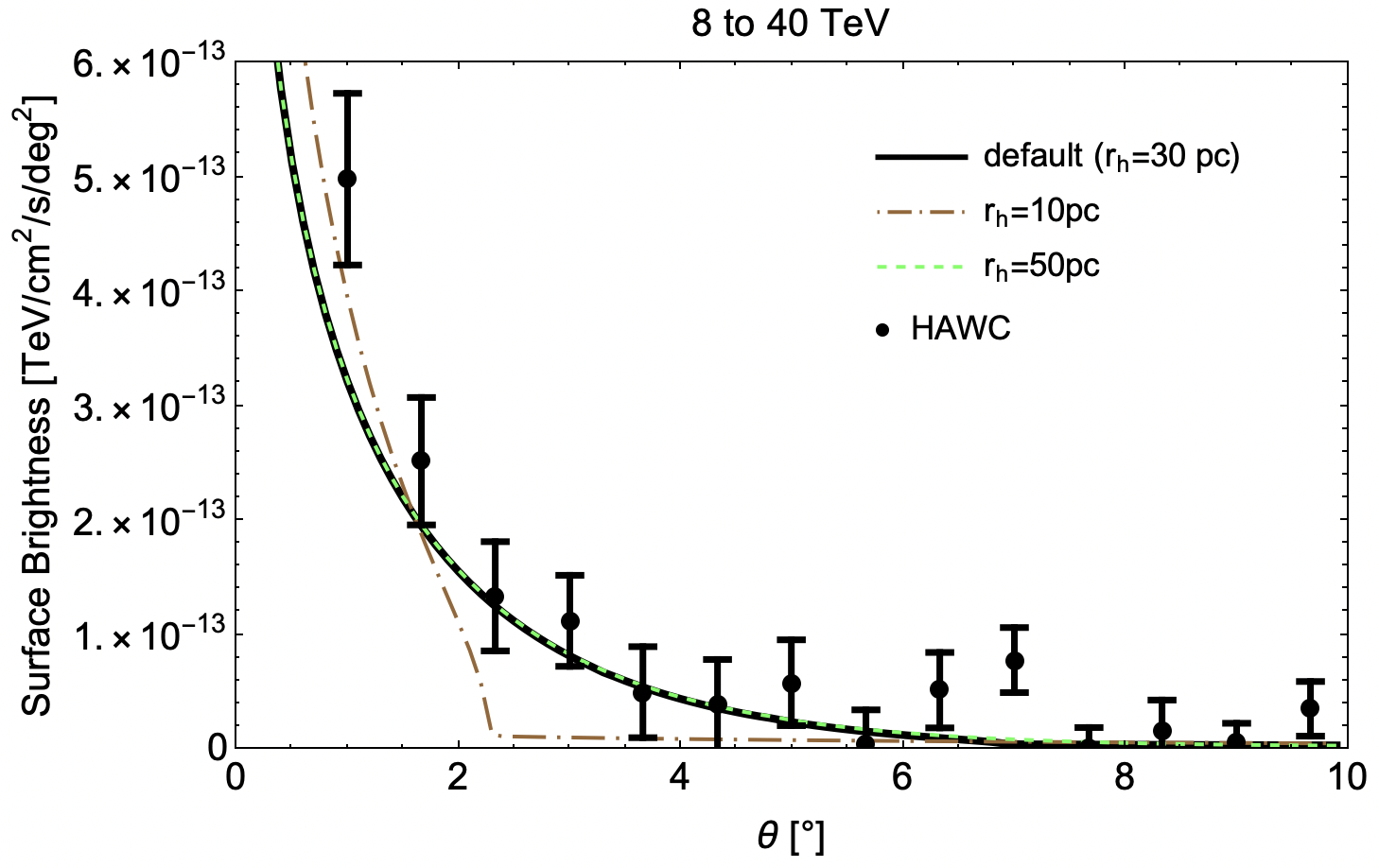}
    \caption{As in Fig.~\ref{fig:SB}, but showing the impact of different parameters in separate frames for clarity.}
    \label{fig:SB_3}
\end{figure}

In Figs.~\ref{fig:SB},~\ref{fig:SB_2}, and~\ref{fig:SB_3}, we illustrate the impact of these same parameters on the surface brightness profile of the gamma-ray emission from a Geminga-like TeV halo. These results are shown as integrated over two ranges of energy: 0.3 to 3 TeV (as could be measured by CTA) and 8 to 40 TeV (as has been measured by HAWC~\cite{HAWC:2017kbo}). From the upper left frames of Figs.~\ref{fig:SB_2} and~\ref{fig:SB_3} we see that the pulsar braking index, $n$, and pulsar spin-down timescale, $\tau$, have little impact on the predicted surface brightness profile. These measurements, however, are much more sensitive to the other parameters considered in this study. Note that some of these parameters can impact the predicted angular distribution in different ways over different ranges of energy, making the results of HAWC and CTA highly complementary. For example, notice that although the different choices of $\alpha$ and $E_{\rm cut}$ considered here only modestly impact the surface brightness profile at HAWC energies, these parameters have a greater impact in the energy range that will be measured by CTA.

\section{The Cherenkov Telescope Array}

The Cherenkov Telescope Array (CTA) will be the flagship of the next-generation instruments in the field of gamma-ray astronomy. It will cover an extensive energy range from 20~GeV to 300~TeV with an energy resolution better than 10\%, and with much greater angular resolution than existing gamma-ray telescopes~\cite{CTAConsortium:2010umy}. CTA will include two different telescope arrays; one in each hemisphere. CTA North will consist of 4 large (23~m diameter) and 9 medium sized (12~m diameter) telescopes. In contrast, CTA South will consist of 4 large and 14 medium sized telescopes, along with 37 smaller (4~m) telescopes.

To assess CTA's ability to distinguish between different models of TeV halos, we have made use of the publicly available code \textit{gammapy}~\cite{Gammapy:2023gvb} and have adopted the \textit{prod5} instrument response function (specifically, \textit{North-20deg-AverageAz-4LSTs09MSTs.180000s-v0.1}). Using this software, we simulated mock data for each model of Geminga's TeV halo, considering a total of 50 hours of observation by CTA North. We have taken CTA's field-of-view to be a $5^{\circ}\times 5^{\circ}$ region centered on Geminga and have divided the data into $0.05^{\circ} \times 0.05^{\circ}$ angular bins, as well as 20 energy bins distributed logarithmically between 0.03 and 100 TeV. Each simulation incorporates both signal and background photons, utilizing the standard background model, \textit{FoVBackgroundModel}, which employs modern models for the isotropic and Galactic diffuse emission~\cite{Fermi-LAT:2019yla}.

\begin{figure}
    \centering
    \includegraphics[width=0.65\linewidth]{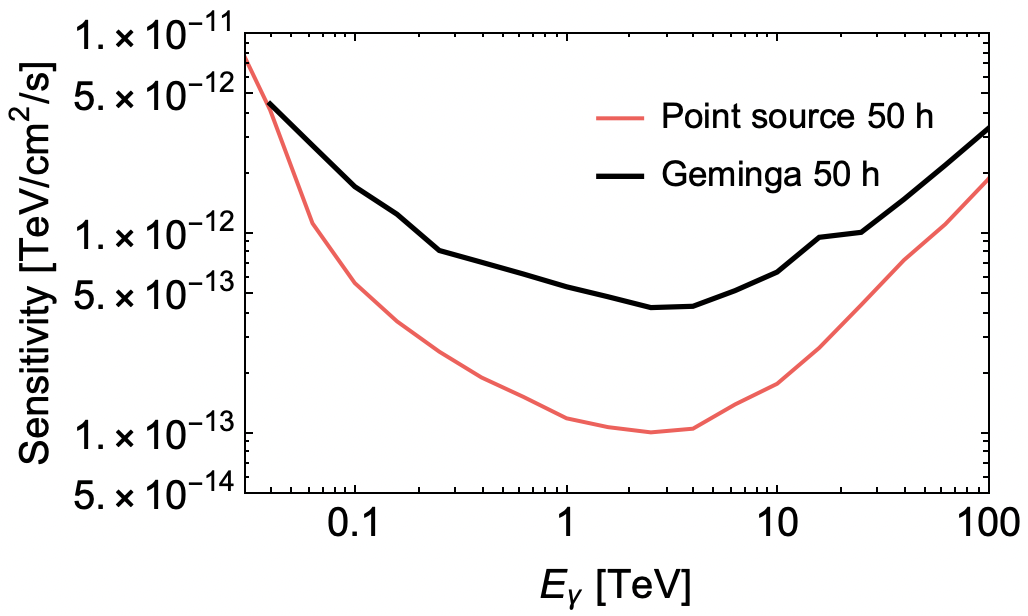}
    \caption{The $5\sigma$ sensitivity of CTA North to a source with an angular extent equal to that of Geminga's TeV halo (adopting our default parameters), after 50 hours of observation. This sensitivity was calculated independently in each of 20 energy bins. For comparison, we also show the analogous sensitivity of CTA to a point source~\cite{CTAperormance}.}
    \label{fig:sensitivity}
\end{figure}

In Fig.~\ref{fig:sensitivity}, we show the $5\sigma$ sensitivity of CTA North to a source with an angular extent equal to that of Geminga's TeV halo (adopting our default parameters), after 50 hours of observation. This sensitivity was calculated independently in each of 20 energy bins. For comparison, we also show the analogous sensitivity of CTA to a point source. The extended nature of Geminga's TeV halo non-negligibly reduces CTA's sensitivity.

\section{Results}

In this section, we project the ability of CTA to test and discriminate between various models of TeV halos, using the Geminga TeV halo as a case study. To this end, we utilize the simulation described in the previous section to calculate the predicted flux in each bin for a given model. We then compute the expected error bars around this flux, $\Delta F = F \times (\Delta N/N_S)$, where $\Delta N = \sqrt{N_{S} + N_{BG}}$, $N_S$ is the predicted number of signal events, and $N_{BG}$ is the predicted number of background events.

\begin{figure}
    \centering
    \includegraphics[width=0.99\linewidth]{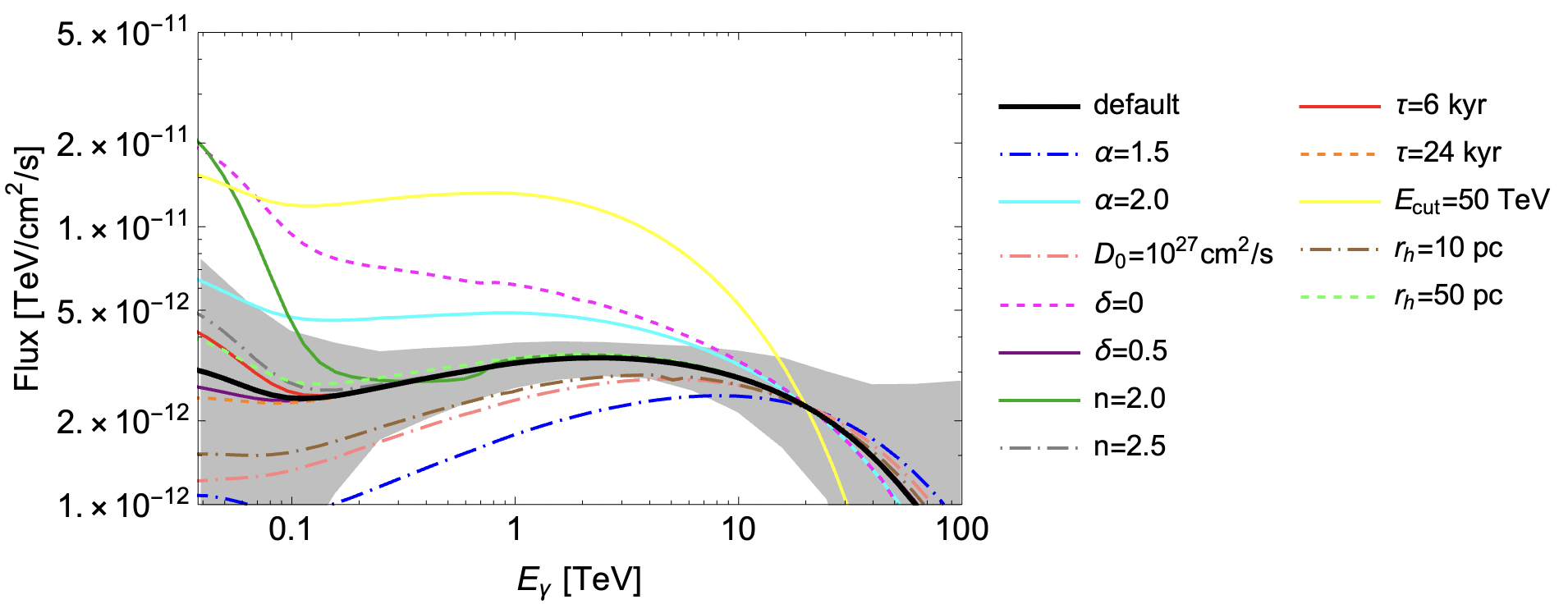}
    \caption{As in Fig.~\ref{fig:flux}, with the shaded grey region representing the projected $1\sigma$ uncertainties after 50 hours of observation of the Geminga TeV halo with CTA North, for the case of our default model.}
    \label{fig:flux_2}
\end{figure}

Our results as they pertain to the energy spectrum of Geminga are shown in Fig.~\ref{fig:flux_2}, where the shaded band reflects our projection for CTA's uncertainties, as calculated (at the $1\sigma$ level) independently in each energy bin, and adopting the case of our default TeV halo model. A casual inspection of this figure reveals that CTA will be able to differentiate our default model of Geminga's TeV halo from models with a relatively low energy cut off ($E_{\rm min}=50 \, {\rm TeV}$), or that feature energy-independent diffusion ($\delta=0$). CTA will also be able to infer the injected spectral index of electrons, easily distinguishing between models with $\alpha=2.0$, 1.8 or 1.5. CTA should also be able to test models with a small halo radius ($r_h=10 \, {\rm pc}$), or with a larger diffusion coefficient than adopted in our default model ($D_0 = 10^{27} \, {\rm cm}^2/{\rm s}$). In contrast, this spectral information will not be able to discriminate our default model from models with $\delta=0.5$, or be very sensitive to the values of $\tau$ or $n$.

\begin{figure}
    \centering
    \includegraphics[width=0.99\linewidth]{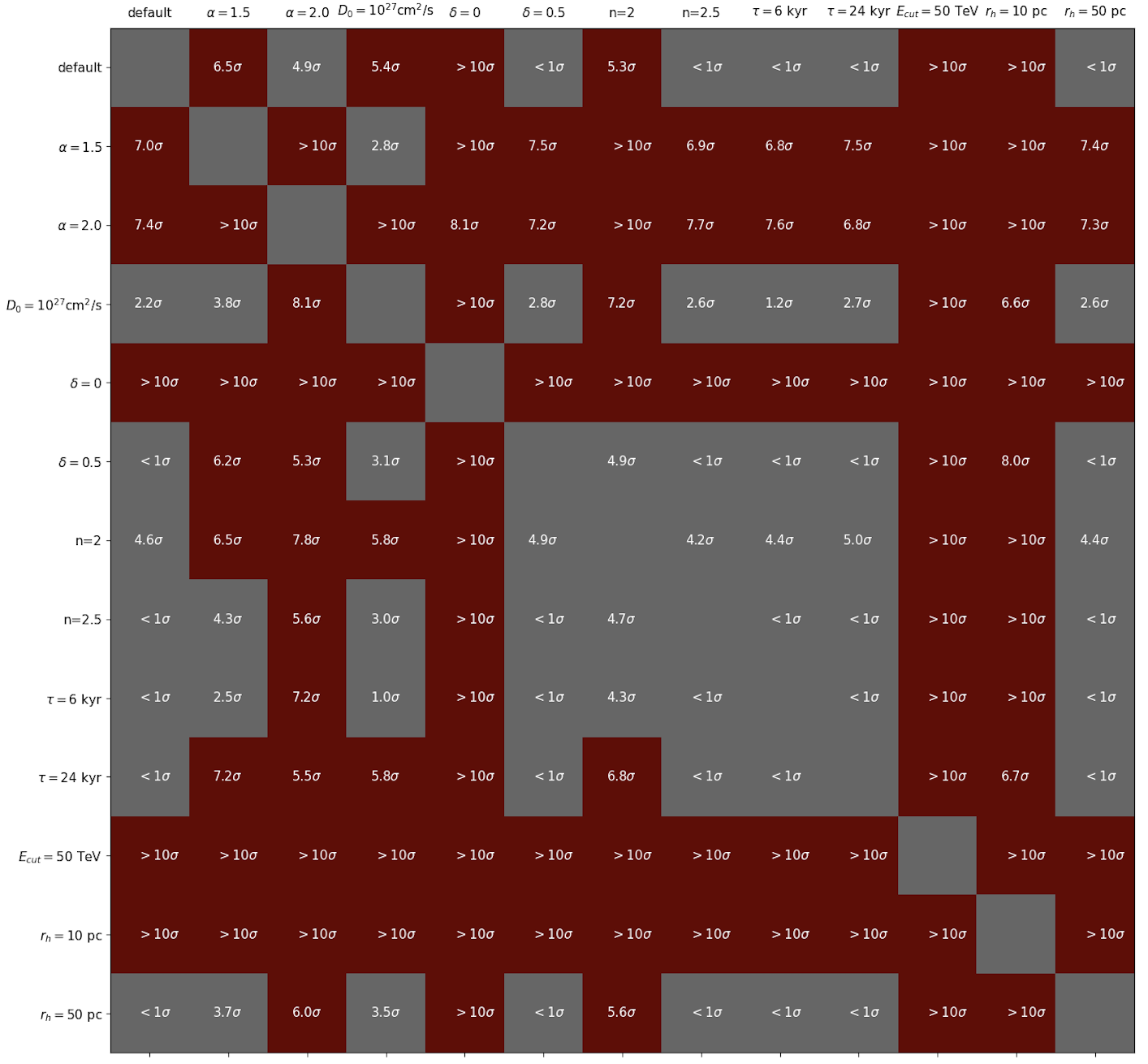}
    \caption{The statistical significance at which CTA is projected to be able to distinguish between different models for Geminga's TeV halo, after 50 hours of observation. Taking a given model to be the ``true model'' of Geminga's TeV halo (shown on the $x$-axis), we calculate the quality of the fit for each of the models considered in this study (shown on the $y$-axis). This information is then converted into the number of standard deviations at which the two models can be distinguished. Models that can be distinguished at the level of 5$\sigma$ or more are shown in red, while other combinations are shown in grey.}
    \label{fig:logL}
\end{figure}

In Fig.~\ref{fig:logL}, we forecast the statistical significance at which CTA will be able to distinguish between different models of Geminga's TeV halo, assuming 50 hours of observation.
Taking a given model to be the ``true model'' of Geminga's TeV halo (shown on the $x$-axis), we calculate the reduced $\chi^2$ of the fit for each of the models considered in this study (shown on the $y$-axis). Each reduced $\chi^2$ is then converted into a $p$-value and then into a statistical significance. Models that can be distinguished at the level of 5$\sigma$ or more are shown in red, while other combinations are shown in grey. Note that these results take into account both spectral and spatial information, allowing us to differentiate between models that the measured spectrum alone would not be able to distinguish.

\section{Summary and Conclusions}

Observations by HAWC, LHAASO, and HESS have revealed the presence of bright and spatially extended multi-TeV emission from the regions surrounding many pulsars, including the nearby examples of Geminga and Monogem. The characteristics of this emission indicate that these sources convert on the order of 10\% of their total spin-down power into very high-energy electrons and positrons which then generate the observed gamma rays through inverse Compton scattering. Surprisingly, the gamma-ray emission from TeV halos is observed to extend out to tens of parsecs in radius, requiring that cosmic rays propagate much less efficiently in the vicinity of these sources than they do elsewhere in the interstellar medium (ISM). How and why diffusion is inhibited within these regions remains an open question. To test and differentiate between different models which could potentially account for these observations will require more detailed measurements of the spectrum and morphology of the gamma-ray emission from these sources.

In this paper, we have studied the ability of the Cherenkov Telescope Array (CTA) to study the properties of TeV halos, focusing on the prototypical example of Geminga. We have considered a variety of models with different values for the parameters associated with the injected electron spectrum, the time evolution of the pulsar's spin-down, and that describe the process of diffusion in the region surrounding the pulsar. We have identified many models that are consistent with all existing data, but that we project could be differentiated by CTA (see Figs.~\ref{fig:flux_2} and~\ref{fig:logL})

\acknowledgments

We would like to thank Ilias Cholis, Luca Orusa, Ievgen Vovk, Igor Moskalenko, Volodymyr Savchenko and Vadym Voitsekhovskyi for helpful discussions. We acknowledge support from {\sl Fermi Research Alliance, LLC} under Contract No. DE-AC02-07CH11359 with the U.S. Department of Energy, Office of High Energy Physics.

\bibliographystyle{JHEP}
\bibliography{ref}

\end{document}